
\magnification=\magstep1
\baselineskip=20 truept
\def\ni{\noindent}

\def\bn{\bigskip\noindent}
\def\sn{\smallskip\noindent}
\def\ie{{\it i.e.}}
\def\pa{\partial}
\def\sech{\mathop{\rm sech}\nolimits}
\def\SN{\mathop{\rm sn}\nolimits}
\def\CN{\mathop{\rm cn}\nolimits}
\def\DN{\mathop{\rm dn}\nolimits}
\def\tr{\mathop{\rm tr}\nolimits}
\def\diag{\mathop{\rm diag}\nolimits}
\def\half{{\textstyle{1\over2}}}
\def\ii{{\rm i}}  
\def\vp{\varphi}

\font\scap=cmcsc10
\font\title=cmbx10 scaled\magstep1

\def\a{\alpha}
\def\b{\beta}
\def\l{\lambda}

\def\c{\gamma}
\def\r{\rho}
\def\t{\theta}
\def\o{\omega}
\def\z{\zeta}
\def\D{\Delta}

\def\ddt{{{\rm d}\over {\rm d}t}}
\rightline{DTP-98/77}
\vskip 1truein
\centerline{\title Two Integrable Systems Related to Hyperbolic Monopoles.}

\vskip 1truein
\centerline{\scap R. S. Ward}

\bn\centerline{\it Dept of Mathematical Sciences, University of Durham,}
\centerline{\it Durham DH1 3LE, UK.}

\vskip 2truein
\ni{\bf Abstract.} Monopoles on hyperbolic 3-space were introduced by
Atiyah in 1984.  This article describes two integrable systems which are
closely related to hyperbolic monopoles: a one-dimensional lattice equation
(the Braam-Austin or discrete Nahm equation),
and a soliton system in (2+1)-dimensional anti-deSitter space-time.

\vskip 1truecm
\ni To appear in:

\ni{\it Asian Journal of Mathematics}, Special 70th Birthday
    Issue for Sir Michael Atiyah.
\vfil\eject

\bn{\scap 1. Introduction.}

\sn Sir Michael Atiyah has made important contributions in several areas of
mathematical physics, and these include the study of
Bogomolny-Prasad-Sommerfield
(BPS) monopoles; see, for example, Atiyah and Hitchin (1979).   Static
BPS monopoles are solutions of a nonlinear elliptic partial
differential equation on some three-dimensional Riemannian manifold.
Most work on monopoles has dealt with the case when this manifold is
Euclidean space  $R^3$: the equations are then completely-integrable, and
can be handled by geometrical techniques.  But the monopole equations on
hyperbolic space $H^3$ are also integrable, as was pointed out by
Atiyah (1984a,b); and in some ways, hyperbolic monopoles are simpler than
Euclidean ones. Hyperbolic monopoles tend to
Euclidean monopoles as the curvature of the hyperbolic space tends to
zero, although this is a delicate fact which was only recently established
(Jarvis and Norbury 1997).

This note describes two integrable systems which are intimately related
to, and were motivated by, hyperbolic monopoles.  The first is a discrete
system (or integrable mapping), which in a certain sense is dual to the
hyperbolic monopole system; this is the discrete Nahm or Braam-Austin
equation.  The second comes from replacing the positive-definite space $H^3$
by a Lorentzian version,
namely anti-deSitter space.  The Bogomolny equation
becomes an evolution equation on this space-time, admitting soliton
solutions.


\bn{\scap 2. Hyperbolic Monopoles and the Discrete Nahm Equations.}

\sn Motivated by the monad construction for
instantons used by Atiyah et al (1978), Nahm (1982) discovered a kind of
duality (subsequently called {\it reciprocity}: see Corrigan and Goddard 1984)
between the monopole equations and solutions of a nonlinear ordinary
differential equation.  This ODE (described below) is called the Nahm
equation.

For monopoles on hyperbolic space $H^3$ of curvature $-C$, a variant of
the Nahm construction works (Braam and Austin 1990), at least 
if $C^{-1}$ is a positive integer.  Such ``integral''
hyperbolic monopoles correspond to certain solutions of a discrete Nahm
equation: a nonlinear difference equation defined on $C^{-1}$ lattice sites.
This is actually a special case of the ADHM construction (Atiyah et al 1978)
for instantons.

One has a picture, therefore, of a correspondence (reciprocity) between
monopoles on the 3-space of constant negative curvature $-C$, and solutions
of the discrete Nahm equation on a one-dimensional lattice with lattice
spacing $C$ (provided $C$ is the inverse of an integer).  The limit $C\to0$
is the continuum limit, in which the discrete Nahm (difference) equation
becomes the Nahm (differential) equation.  It seems likely on general grounds
that reciprocity
operates, in some sense, for non-integral hyperbolic monopoles; but this
remains an open question.
In what follows, I shall concentrate just on the discrete Nahm equations,
and not say anything about hyperbolic monopoles.

Let $k$ be a fixed positive integer; and let $A_j$,  $B_j$,
$C_j$ and  $D_j$ denote $k\times k$ matrices, defined for each value of the
integer $j$. In other words, we have four  $k\times k$ matrices on a
one-dimensional lattice indexed by $j$.  Let $s_{\pm}$ denote the forward
and backward step
operators on this lattice: so $(s_+ \Psi)_j = \Psi_{j+1}$ and
$(s_- \Psi)_j = \Psi_{j-1}$.  For brevity, the subscript $j$ will usually be
omitted in what follows; thus $A$ stands for $A_j$,
$A_+ = s_+ A$ stands for $A_{j+1}$, and so forth.

Consider the two linear operators
$$
\eqalign{ U &:= C s_+ + \l A \cr
          V &:= D_- s_- - \l^{-1}B \cr}
$$
(acting on a $k$-vector $\Psi$ defined on the lattice). Here $\l$ is a constant
scalar parameter, and the minus signs are mainly for notational convenience.
The eigenvalue equations
$$\eqalign{
  U\Psi &:= C\Psi_+ + \l A\Psi = \zeta\Psi, \cr
  V\Psi &:= D_-\Psi_- -\l^{-1}B\Psi = \mu^{-1}\Psi \cr} \eqno(1)
$$
are difference equations
which propagate  $\Psi$ forwards and backwards along the lattice.
In order for a nontrivial solution (simultaneous eigenfunction of $U$ and $V$)
to exist, we need $U$ and $V$ to commute, and the parameters $\l$, $\zeta$,
$\mu$ to satisfy an algebraic relation (which turns out to be the vanishing
of a homogeneous polynomial in these three variables).
The condition $[U,V]=0$ gives the discrete Nahm equation; 
equation (1) is a Lax pair for it; and the algebraic relation defines a
spectral curve (from which one can derive conserved quantities, show
that the solutions of discrete Nahm correspond to stepping along straight lines
on the Jacobian of this curve, etc: see Murray and Singer 1998).

The condition that $[U,V]=0$ should hold for all $\l$ is equivalent to
$$
\eqalign{ A_+ &= D A D^{-1}, \cr
          B_+ &= C^{-1} B C, \cr
          C_+ D_+ &= DC + [A_+, B_+]. \cr } \eqno(2)
$$
These are the discrete Nahm (or Braam-Austin) equations.
They consist of three difference equations for four (matrix) functions:
the under-determinacy reflects the gauge freedom in (2). Namely, if
$\Lambda$ is a non-singular matrix on the lattice, then the system (2)
is invariant under the gauge transformations
$$
\eqalign{ A &\mapsto \Lambda A \Lambda^{-1} \cr
          B &\mapsto \Lambda B \Lambda^{-1} \cr
          C &\mapsto \Lambda C \Lambda_+^{-1} \cr
          D &\mapsto \Lambda_+ D \Lambda^{-1}. \cr} \eqno(3)
$$
A gauge choice such as $D=C$ converts (2) into a determined system: three
matrix difference equations for three matrices.

For completeness, let us note the correspondence with the notation of
Braam and Austin (1990), where the equations (2) were first derived.
The relation between $A$, $B$, $C$,
$D$ and their matrices $\beta_j$, $\gamma_j$ is given by $A=B^*$,
$D=C^*$, and
$$
\eqalign{\beta_{2j} &= B_j\,, \cr
         \gamma_{2j+1} &= C_j\,. \cr}
$$

I shall not describe the spectral curve and its consequences here; but merely
list the eight independent conserved quantities  in the $k=2$ case.
They are $\tr A$,  $\tr A^2$, $\tr B$,  $\tr B^2$,
$\tr(CD)$,  $\tr(ACD)$, $\tr(BCD)$,  and $\tr[CD(CD-2AB)]$.  Each of these 
expressions is constant on the lattice, by virtue of (2); note that they are
also gauge-invariant.

A continuum limit of (2) may be obtained as follows. Replace the integer
variable $j$ by $t=jh$, where $h$ is the ``lattice spacing'', and take the
limit $h\to0$.  Write
$$
\eqalign{ C &= (2h)^{-1} I + \half\ii T_3 = D, \cr
          B &= \half(T_1 + \ii T_2) = - A^*,\cr } \eqno(4)
$$
where the $T_\a$ are antihermitian $k\times k$ matrices,
$I$ denotes the identity matrix,
and star denotes complex conjugate transpose.  Then the  $h\to0$ limit of (2)
is
$$
  \ddt T_\a + \half\varepsilon_{\a\b\c}\bigl[ T_\b , T_\c \bigr] = 0, \eqno(5)
$$
which are the Nahm equations.
Similarly, one obtains the standard Lax pair for the Nahm equations as
the continuum limit of (1), if $\zeta$ and $\mu$ are chosen appropriately.

\bn{\scap 3. Reduction to a Discrete Toda System.}

\sn It has long been known that the Nahm equation reduces to the Toda
lattice.  The $T_\a$ take values in a Lie algebra, which (in the simplest
case that we are considering here) is $su(k)$.  To reduce to Toda, one
takes $T_3$ in a Cartan subalgebra, and $T_1 \pm \ii T_2$ corresponding
to $\pm$ a set of simple roots.  What follows is the discrete version of this
reduction.

We express $A$, $B$, $C$ and $D$ in terms of $2k$ lattice functions
$f_a = f_{aj}$, $p_a = p_{aj}$, where $a = 1,2,\ldots,k$, as follows:
$C = D = \diag(f_1,f_2,\ldots,f_k)$ and
$$
B = \pmatrix{0 &p_2 &0   &\ldots &0 \cr
             0 &0   &p_3 &\ldots &0 \cr
             \vdots&\vdots&\vdots&\ddots&\vdots \cr
             0 &0 &0 &\ldots &p_k \cr
             p_1 &0 &0 &\ldots &0 \cr} = -A^*\,. \eqno(6)
$$
Then the discrete Nahm equations (2) reduce to
$$\eqalign{
 (p_a)_+ &= p_a f_a/f_{a-1}\,,\cr
 f_a^2 &= (f_a^2)_- + p^2_{a+1} -  p^2_a\,,\cr} \eqno(7)
$$
where $f_0$ is interpreted as $f_k$, and $p_{k+1}$ as $p_1$ (in other
words, the index $a$ is periodic with period $k$).  The equations (7)
constitute a discrete-time Toda lattice.  The first example of such a
system was that of Hirota (1977), and many other examples have been
described more recently.

Notice that the quantity $\Sigma = \sum_{a=1}^k f_a^2$ is conserved;
in other words, $\Sigma_+ = \Sigma$.  Let us define a parameter $h$ by
$\Sigma = k/h^2$, and assume that $f_a h \to 1$ as $h\to0$, for each $a$.
Then the  $h\to0$ limit of (7) is the differential equation
$$
 {{\rm d}^2\over{\rm d}t^2} \log p_a^2 = p_{a+1}^2 - 2p_a^2 +  p_{a-1}^2,
    \eqno(8)
$$
which is the Toda lattice.

Let us look in detail at the $k=2$ case.  Rewrite $f_a$ and $p_a$ in terms
of three functions $u$, $v$, $w$, and the constant $h$, according to
$$\eqalign{
f_1 &= h^{-1}\sqrt{1+2hw}, \cr
f_2 &= h^{-1}\sqrt{1-2hw}, \cr
p_1 &= u-v, \cr
p_2 &= u+v. \cr
}$$
Then (7) is
$$
\eqalign{w_+ &= w + 2h u_+ v_+ \cr
         u_+ &= (u - 2hwv)/\sqrt{1-4h^2 w^2}  \cr
         v_+ &= (v - 2hwu)/\sqrt{1-4h^2 w^2},  \cr} \eqno(9)
$$
which are a discrete-time version of Euler's equations for a spinning top
(with an appropriate choice of moments of inertia).  Indeed, 
in the continuum limit $h\to0$, (9) becomes ${\rm d}w/{\rm d}t = 2uv$,
${\rm d}u/{\rm d}t = -2vw$, ${\rm d}v/{\rm d}t = -2wu$.
These are Euler's equations, which can
be solved in terms of elliptic functions; and (9) can be solved
likewise, as follows.

Notice that (9) admits two independent conserved quantities, namely
$$
\eqalign{\Theta &= u^2 - v^2, \cr
         \Omega &= 2w^2 + u^2 + v^2 - 4hwuv. \cr}  \eqno(10)
$$
This enables one to express $u$ and $v$ in terms of $w$ (and $\Theta$,
$\Omega$); and then the first equation in (9) leads to a difference equation
for $w$ alone.  Its solution is
$$
  w_j = {k\over2h} \SN(bh) \SN(bjh + c),  \eqno(11)
$$
where $k$, $b$ and $c$ are ``constants of integration''.  Here $k$ denotes
the modulus of the elliptic functions.  The conserved quantities $\Theta$
and $\Omega$ are related to $k$ and $b$ as follows:
$$\eqalign{
\Theta &= (2h^2)^{-1} \bigl[ \CN(bh) - \DN(bh) \bigr], \cr
\Omega &= (2h^2)^{-1} \bigl[ 1 - \CN(bh)\DN(bh) \bigr]. \cr
}$$
The functions $u$ and $v$ are then given by
$$\eqalign{
(u+v)^2 &= (\Omega -2w^2 + \D)/(1 - 2hw), \cr
(u-v)^2 &= (\Omega -2w^2 - \D)/(1 + 2hw), \cr} \eqno(12)
$$
where $\D_j = h^{-1} \SN(bh) \CN(bjh + c) \DN(bjh + c)$.  To see that
$u$ and $v$ are well-defined (and real) for all $b$, $c$ and $0<k<1$,
note first that $1\pm 2hw$ is positive.  Secondly, $\Omega - 2w^2 \geq 0$,
from the inequality $1 - \CN \DN \geq k^2 \SN^2$.  Finally,
$(\Omega-2w^2)^2 - \D^2 = (1-4h^2w^2)\Theta^2 \geq 0$.
As a final remark, note that in the limiting case $k=1$, the solution is
$$\eqalign{
w    &= (2h)^{-1} \tanh(bh) \tanh(bt), \cr
v = u&= (2h)^{-1} \tanh(bh) \sech(bt) \bigr/ \sqrt{1+\tanh(bh) \tanh(bt)}. \cr
}$$

It seems likely that this elliptic $k=2$ solution corresponds to hyperbolic
2-monopoles with gauge group SU(2), via the Braam-Austin construction.
More generally, for $k>2$ one may speculate that discrete-Toda solutions
correspond to hyperbolic $k$-monopoles with $C_k$ cyclic symmetry, since
this is what happens for Euclidean monopoles (Sutcliffe 1996).


\bn{\scap 4. Solitons in (2+1)-Dimensional Anti-deSitter Space-Time.}

\sn Let $M$ be a three-dimensional Riemannian manifold, with metric
$g$ and volume element $\eta$.  We are interested in Yang-Mills-Higgs fields
on $M$, with gauge group SU(2) (for simplicity).   So we have a Higgs field
$\Phi = \Phi(x^\mu)$ taking values in the Lie algebra su(2); here
$x^\mu = (x^0,x^1,x^2)$ are local coordinates on $M$.  A gauge potential
(connection) $A_\mu(x^\nu)$ determines the covariant derivative
$D_\mu \Phi = \pa\Phi/\pa x^\mu + [A_\mu , \Phi]$.  The gauge field
(curvature) is the su(2)-valued 2-form $F_{\mu\nu} = [D_\mu , D_\nu]$.
And the Bogomolny equations for $(A_\mu , \Phi)$ are
$$
   D \Phi = *F, \eqno(13)
$$
or, in index notation,
$$
   D_\mu \Phi = \half g_{\mu\nu} \eta^{\nu\a\b} F_{\a\b}. \eqno(13')
$$
These are coupled nonlinear partial differential equations which, in general,
are not completely integrable.  But (13) {\it is} an integrable system (in
the sense
that a Lax pair exists) if the metric $g$ has constant curvature.  For
example, if $(M,g)$ is Euclidean space $R^3$ or hyperbolic space $H^3$, then
(13) is the equation for Euclidean or hyperbolic BPS monopoles, respectively.

Another possibility is for $g$ to have Lorentzian signature $-++$,
and then (13) are evolution equations in the space-time $(M,g)$.
Soliton solutions in the case of flat space-time have been studied
in some detail: see Ward (1988, 1990, 1998).  The aim here is
to describe an example in curved space-time.

There are two curved space-times with constant curvature: deSitter space
with positive scalar curvature $R$, and anti-deSitter space with $R < 0$
(Hawking and Ellis 1973).  I shall deal here with the latter case only,
namely anti-deSitter space (AdS).  By definition, (2+1)-dimensional
anti-deSitter space is the universal
covering space of the hyperboloid ${\cal H}$ with equation
$$
  U^2 + V^2 - X^2 - Y^2 = 1, \eqno(14)
$$
and with metric induced from
$$
  ds^2 = -dU^2 - dV^2 + dX^2 + dY^2. \eqno(15)
$$
If, for example, we parametrize the  hyperboloid ${\cal H}$ by
$$\eqalign{
U &= \sec\r \cos\t \cr
V &= \sec\r \sin\t \cr
X &= \tan\r \cos\vp \cr
Y &= \tan\r \sin\vp \cr
} \eqno(16)
$$
with $0\leq \r < \pi/2$, then we get the metric
$$
  ds^2 = \sec^2\r (-d\t^2 + d\r^2 + \sin^2\r\, d\vp^2 ). \eqno(17)
$$
At this stage, the space-time contains closed timelike curves, because of
the periodicity of $\t$.  Anti-deSitter space is the universal cover of
${\cal H}$, in which $\t$ is unwound (so that $\t\in R$).
Consequently, AdS, as a manifold, is the product of an open spatial disc
(on which $\r$
and $\vp$ are polar coordinates) with time  $\t\in R$.   It is a space
of constant curvature, with scalar curvature equal to $-6$.  Null/spacelike
infinity ${\cal I}$ consists of the timelike cylinder $\r = \pi/2$; this
surface is never reached by timelike geodesics.

In what follows below, we shall also use Poincar\'e coordinates $t$,
$x$ and $r > 0$.  They are defined by
$$\eqalign{
t &= -V/(U+X) \cr
r &= 1/(U+X) \cr
x &= Y/(U+X), \cr
} \eqno(18)
$$
in terms of which the metric is 
$$
  ds^2 = r^{-2} (-dt^2 + dr^2 + dx^2 ). \eqno(19)
$$
But these only cover a small part of AdS, corresponding to half
$U+X>0$ of the hyperboloid ${\cal H}$.  The surface $r=0$ is part of
infinity ${\cal I}$.

The minitwistor space corresponding to AdS, or rather to the Poincar\'e
space (19), is $CP^1\times CP^1$, which we visualize as a quadric $Q$ in $CP^3$
({\it cf.}\ Hitchin 1982).  The points of space-time correspond to certain
plane sections (conics) of $Q$.  The space of all planes is a $CP^3$.
But the relevant conics have to be real (which in this case means that
their defining planes have real coefficients), and nondegenerate.  So the
space of these acceptable conics is the ``top half'' of $RP^3$,
parametrized by the real homogeneous coordinates $(U,V,X,Y)$ with
$U^2 + V^2 - X^2 - Y^2 > 0$.   This $RP^3_+$ is double-covered by the
original hyperboloid ${\cal H}$, and is essentially the Poincar\'e space
(19).   More accurately, the coordinates $(t,r,x)$ cover all of $RP^3_+$
except for a set of measure zero.  If $\o$ and $\z$ are standard coordinates
on the two $CP^1$ factors of $Q$, then the conics are
$$
 \o = \o(\z) = {v\z - (uv+r^2) \over \z - u}
            = {(Y+V)\z+(X-U) \over (U+X)\z+(V-Y)}, \eqno(20)
$$
where $u = x+t$ and $v = x-t$.  Eqn (20) expresses the correspondence
between space-time and twistor space $Q$.

The idea now is that holomorphic vector bundles $V$ over $Q$ (saisfying some
mild conditions) determine multi-soliton solutions of (13) in anti-deSitter
space, via the usual Penrose transform.
In our case, the relevant vector bundles are stable bundles of rank 2,
with Chern numbers $c_1 = 0$ and $c_2 = 2n$, $n$ being a positive integer.
In the simplest case $n=1$, the moduli space of such bundles is
5-complex-dimensional
({\it cf.}\ Hurtubise 1986, Buchdahl 1987).  When we impose reality conditions,
which amounts to taking the gauge group to be SU(2) rather than SL(2,C),
the moduli space becomes 5-real-dimensional.  So we expect, in this simplest
case, to get a five-parameter family of soliton solutions, exactly as for
the flat-space-time system (Ward 1988, 1990, 1998).   This is exactly what
happens (Hickin 1998).

One explicit way of seeing how solutions arise is as follows: it involves a Lax
pair for the integrable system (13).  Define two operators $\nabla_1$
and  $\nabla_2$ by
$$\eqalign{
 \nabla_1 &= r\pa_r - 2(\z - u)\pa_u \cr
 \nabla_2 &= 2\pa_v + r^{-1}(\z - u)\pa_r. \cr
}\eqno(21)
$$
Notice that  $\nabla_1$ and  $\nabla_2$ both annihilate the expression (20).
This is related to the fact that twistor space $Q$ is the quotient of
the distribution $\{\nabla_1, \nabla_2\}$ (on the four-dimensional
correspondence space
whose local coordinates are $(t,r,x,\z)$).  The Lax pair involves the
gauge-covariant version of (21), and consists of the pair of equations
$$\eqalign{
 \bigl[ rD_r + \Phi - 2(\z - u)D_u \bigr]\psi &= 0 \cr
 \bigl[ 2D_v + r^{-1}(\z - u)(D_r - r^{-1}\Phi) \bigr]\psi &= 0, \cr
}\eqno(22)
$$
where $\psi = \psi(t,r,x,\z)$ is a $2\times2$ matrix.  The consistency
condition for this overdetermined system is exactly (13).

The functions $\psi$ corresponding to $n=1$ bundles can be taken to have
the rational form
$$
  \psi = I - {(\bar\z_0 - \z_0) \over (\z - \z_0)}
               {p^*\otimes p \over p\cdot p^*},  \eqno(23)
$$
where $I$ denotes the identity $2\times2$ matrix, $\z_0$ is a complex constant,
$p(t,r,x)$ is a row 2-vector of linear functions of $\o_0 = \o(\z_0)$,
and $p^*$ denotes its complex conjugate transpose.  So $p$ has the form
$$
   p = (a\o_0 + b, c\o_0 + d),  \eqno(24)
$$
where $a$, $b$, $c$, $d$ are complex constants with $ad-bc\neq0$.  The
Yang-Mills-Higgs fields $(\Phi, A_\mu)$ can then be read off from (22--24),
and they will automatically satisfy (13).  The parameters $\z_0$,
$a$, $b$, $c$, $d$ are not all significant: it is clear
from (23) that an overall complex scaling of $p$ will not change $\psi$,
and furthermore that multiplying $p$ on the right by a constant SU(2) matrix
will induce a gauge transformation on $(\Phi, A_\mu)$.  Removing this freedom
leaves us with a five-real-parameter family of solutions.

Each of
these solutions represents a single soliton (lump), and the five parameters
describe the location (2), velocity (2) and size (1) of this soliton.
It is straightforward
to write down the fields  $(\Phi, A_\mu)$ explicitly as rational
functions of  $t$, $r$, $x$ (or $U$, $V$, $X$, $Y$) --- but these expressions
are not immediately transparent, and we shall make do with the following
remarks.  The solitons are
spatially localized, in the sense that $\Phi\to0$ and $F_{\mu\nu}\to0$
as $r\to0$, \ie. as one approaches null/spacelike infinity ${\cal I}$.
To see a simple example, one may set $p = (\o_0, 1)$.  Then 
the positive-definite gauge-invariant quantity $-\tr\Phi^2$ (which is
a good one for visualizing the field) is given by
$$
-\tr\Phi^2 = 8 r^4 \bigl/ \bigl[ (r^2+x^2-t^2)^2+2x^2+2t^2 \bigr]^2.
              \eqno(25)
$$
The graph of this function is
a single lump, with its maximum along the timelike geodesic
$x=0$, $r^2 = t^2 + 1$: a soliton in free fall.


\bn{\scap 5. Concluding Remarks.}

\sn Many of the ramifications of Atiyah's (1984b) work on hyperbolic monopoles
are only now being addressed.  In the positive-definite case, a study of the
relation between hyperbolic monopoles (and their symmetries) and the
corresponding spectral curves is currently underway (Murray and Singer 1998).
In the Lorentzian case, soliton solutions and their corresponding
vector bundles are being investigated (Hickin 1998); specific questions
include multi-soliton ($n>1$) dynamics, and what happens in deSitter
(rather than anti-deSitter) space.


\bn{\scap Acknowledgements.}

\sn I am grateful to Michael Singer for telling me about his and M K Murray's
work prior to its publication; and to the referee for emphasizing the
possible relevance of the material in Section 3 to symmetric hyperbolic
monopoles.

\bn{\scap References.}

\item{$\bullet$} Atiyah M F 1984a {\it Commun.\ Math.\ Phys.} {\bf93}, 437.
\item{$\bullet$} Atiyah M F 1984b Magnetic monopoles in hyperbolic space. In:
        {\it Proc.\ Bombay Colloquium on Vector Bundles} (Tata Institute
        and Oxford University Press).
\item{$\bullet$} Atiyah M F, Drinfeld V G, Hitchin N J and Manin Yu I 1978
        {\it Phys.\ Lett.\ A} {\bf65}, 185.
\item{$\bullet$} Atiyah M F and Hitchin N J 1979 {\it The Geometry and
        Dynamics of Magnetic Monopoles} (Princeton University Press).
\item{$\bullet$} Braam P and Austin D M 1990 {\it Nonlinearity} {\bf3}, 809.
\item{$\bullet$} Buchdahl N P 1987 {\it Math.\ Z.} {\bf194}, 143.
\item{$\bullet$} Corrigan E F and Goddard P 1984 {\it Ann.\ Phys.}
        {\bf154}, 253.
\item{$\bullet$} Hawking S W and Ellis G F R 1973 {\it The Large Scale
        Structure of Space-Time} (Cambridge University Press).
\item{$\bullet$} Hickin D G 1998 personal communication.
\item{$\bullet$} Hitchin N J 1982 Complex manifolds and Einstein's equations.
        In: {\it Twistor Geometry and Non-Linear Systems}, eds
        H D Doebner and T D Palev (Lecture Notes in Mathematics 970,
        Springer-Verlag).
\item{$\bullet$} Hirota R 1977 {\it J.\ Phys.\ Soc.\ Japan} {\bf43}, 2074.
\item{$\bullet$} Hurtubise J 1986 {\it Commun.\ Math.\ Phys.} {\bf105}, 107.
\item{$\bullet$}  Jarvis S and Norbury P 1997 {\it Bull.\ Lond.\ Math.\ Soc.}
         {\bf29}, 737.
\item{$\bullet$} Murray M K and Singer M A 1998 On the complete integrability
       of the discrete Nahm equations. (Draft Edinburgh preprint).
\item{$\bullet$} Nahm W 1982 The construction of all self-dual multimonopoles
        by the ADHM method. In: {\it Monopoles in Quantum Field Theory}, eds
        N S Craigie, P Goddard and W Nahm (World Scientific, Singapore).
\item{$\bullet$} Sutcliffe PM  1996 {\it Phys.\ Lett.\ B} {\bf381}, 129.
\item{$\bullet$} Ward R S 1988 {\it J.\ Math.\ Phys.} {\bf29}, 386.
\item{$\bullet$} Ward R S 1990 {\it Commun.\ Math.\ Phys.} {\bf128}, 319.
\item{$\bullet$} Ward R S 1998 Twistors, geometry and integrable systems.
        In: {\it The Geometric Universe}, eds S A Huggett et al
        (Oxford University Press).

\bye